\documentclass[a4paper, aps, prd, twocolumn, showpacs, superscriptaddress, groupedaddress, preprintnumbers, nofootinbib]{revtex4-1} 
\usepackage{graphicx}     
\usepackage{amsmath, amssymb, amscd, latexsym, bm, braket}   
\usepackage{url}

\begin{document}
\title{Hadron Masses in Strong Magnetic Fields}
\preprint{UT-KOMABA/14-8}
\author{Hidetoshi Taya}
\email{{\tt h_taya@hep1.c.u-tokyo.ac.jp}}
\affiliation{Institute of Physics, The University of Tokyo, Komaba 3-8-1, Tokyo 153-8902, Japan}
\affiliation{Department of Physics, The University of Tokyo, Hongo 7-3-1, Tokyo 113-0033, Japan}

\date{\today}

\begin{abstract}
Hadron masses under strong magnetic fields are studied.  In the presence of strong magnetic fields exceeding the QCD energy scale $ eB \gg \Lambda^2_{\rm QCD} $, ${\rm SU(3)}_{\rm flavor} \otimes {\rm SU(2)}_{\rm spin}$ symmetry of hadrons is explicitly broken so that the quark components of hadrons differ from those with zero or weak magnetic fields $ eB \lesssim \Lambda^2_{\rm QCD} $.  Also, squeezing of hadrons by strong magnetic fields affects the hadron mass spectrum.  We develop a quark model which appropriately incorporates these features and analytically calculate various hadron masses including mesons, baryons and those with strangeness.  
\end{abstract}

\maketitle

\section{Introduction}
Quantum Chromodynamics (QCD) exhibits highly non-trivial behaviors in the presence of strong magnetic fields exceeding the QCD energy scale $eB \gg \Lambda^2_{\rm QCD}$ \cite{kaz13}.  Indeed, lattice simulations are performed without suffering from the notorious sign problem and many interesting phenomena are observed such as (inverse-)magnetic catalysis \cite{eli11, bal12prd, bal12jhep, bru13}, anisotropy in the quark-antiquark potential \cite{cla14} and non-trivial $eB$-dependence in hadron masses \cite{hid13, lus14a, lus14b}, albeit their physical interpretation is still in intense discussions.  On the experimental side, such strong magnetic fields are realized in the peripheral collisions of relativistic heavy ions and possibly in the interior of neutron stars.  Thus, it is also phenomenologically important to improve our understanding of QCD under strong magnetic fields.

In this paper, we discuss hadron masses under strong magnetic fields.  Recent lattice studies \cite{hid13, lus14a, lus14b} calculated $\pi, \rho$ meson masses and observed that (i) not only charged meson masses $M_{\pi^+}, M_{\rho^+}$ but also neutral meson masses $M_{\pi^0}, M_{\rho^0}$ depend on the strength of magnetic fields $eB$; (ii) $M_{\pi^+}, M_{\rho^0}$ increase as $\sqrt{eB}$ while $M_{\pi^0}, M_{ \rho^+}$ show weak $eB$-dependence; (iii) there is a mass hierarchy $M_{\rho^0} \sim M_{\pi^+} > M_{ \rho^+} \sim M_{\pi^0}$ for $eB \gg \Lambda_{\rm QCD}^2$.  So far two phenomenological studies employing a relativistic Hamiltonian technique \cite{and13} and the Nambu-Jona-Lasinio model \cite{hao14} have been done to explain these non-trivial behaviors.  However, there is no consensus on the physical reason why there is such non-trivial behaviors.  

The purpose of this paper is to present a simple analytical model which clearly explains the physics of hadron masses under strong magnetic fields.  We also apply the model to predict other hadron masses including baryons and hadrons with strangeness.

\section{Model Description}
 
Before presenting the model, let us first clarify what are the essential properties for describing hadron masses under strong magnetic fields.

(a) {\it Quark degrees of freedom}: Since the typical energy scale of the system is now characterized by magnetic fields stronger than the QCD energy scale $eB \gg \Lambda^2_{\rm QCD}$, the internal structure of hadrons, i.e., the quark degrees of freedom should be explicitly treated to describe hadron masses.  

(b) {\it Explicit breaking of ${\rm SU(3)}_{\rm flavor} \otimes {\rm SU(2)}_{\rm spin}$ symmetry of hadrons}: ${\rm SU(3)}_{\rm flavor} \otimes {\rm SU(2)}_{\rm spin}$ symmetry of hadrons is the key concept in describing hadron masses with zero magnetic field \cite{gel64, zwe64, ruj75}.  However, this ${\rm SU(3)}_{\rm flavor} \otimes {\rm SU(2)}_{\rm spin}$ symmetry is completely broken under strong magnetic fields because quarks form the Landau levels and their lowest energies depend on the spin and the electric charge.  As a result, the quark components of low-lying hadrons under strong magnetic fields differ from those with zero or weak magnetic fields $ eB \lesssim \Lambda^2_{\rm QCD} $.  

(c) {\it Strong deformation of hadrons}: Without strong magnetic fields, the typical volume of a hadron is solely determined by the confinement of QCD and is roughly given by $(1/\Lambda_{\rm QCD})^3$.  Under strong magnetic fields, however, hadrons are strongly squeezed in the transverse direction $\braket{r} \sim 1/\sqrt{eB}$ not by QCD but by strong magnetic fields and therefore the typical volume would be given by $(1/\sqrt{eB})^2\times (1/\Lambda_{\rm QCD})$.  As a result, the typical distance between quarks inside a hadron $|{\bm r}_{\rm qq}|$ decreases as the magnetic field gets stronger.  Due to this reduction of $|{\bm r}_{\rm qq}|$, the mass contribution from the long-range interaction between quarks, i.e., the confinement potential of QCD $\propto |{\bm r}_{\rm qq}|$ decreases.

When describing hadron masses under strong magnetic fields, one should take into account all of the essential properties (a)-(c).  We are going to develop a simple analytical model which incorporates all of the properties: A quark model under strong magnetic fields (a) whose quark components of hadrons are reorganized to respect the explicit breaking of the ${\rm SU(3)}_{\rm flavor} \otimes {\rm SU(2)}_{\rm spin}$ symmetry (b) and which includes the confinement potential of QCD (c).

Let us consider a Hamiltonian $H$, for a single quark with an electric charge $q$ and a current quark mass $m$, of the form
\begin{eqnarray}
	H ({\bm r})			&=& {\bm \alpha} \cdot (-i \nabla - q {\bm A}({\bm r})) + \beta V({\bm r}), \nonumber\\
	{\bm A}({\bm r}) 	&=& \frac{1}{2} B r {\bm e}_{\theta}, \nonumber\\
	V({\bm r}) 			&=& \sqrt{ m^2 + \sigma_{\perp }^2 r^2 + \sigma_{\parallel}^2 z^2}.  \label{eq3}
\end{eqnarray}
Here, we take cylindrical coordinates $r=\sqrt{x^2+y^2}, \theta = \arctan(y/x)$ and denote the Dirac matrices with ${\bm \alpha}$ and $\beta$.  The vector potential ${\bm A}$ is chosen to realize a constant magnetic field along the $z$-axis ${\bm B} = B{\bm e}_z$.  The potential $V$ contains not only the mass term $m^2$ but also the term $\sigma_{\perp}^2 r^2 + \sigma_{\parallel}^2 z^2$, which phenomenologically represents the linear confinement of QCD.  The parameters $\sigma_{\perp},\sigma_{\parallel}$ characterize the confinement force of QCD\footnote{Although the parameters $\sigma_{\perp}, \sigma_{\parallel}$ resemble to the phenomenological string tension of QCD $\sigma_{\rm ph}$, they are not the same in a strict sense: The phenomenological string tension $\sigma_{\rm ph}$ represents the force acting between a pair of dynamical quarks.  Our parameters $\sigma_{\perp}, \sigma_{\parallel}$ characterize the force acting on a single quark moving in an effective single-particle potential of the confinement.  } in the transverse and the longitudinal direction with respect to the magnetic field, respectively.  Notice that the chiral symmetry is explicitly broken by the potential $V$.  We also stress that the detailed choice of the confinement potential in $V$ does not affect our qualitative results.  The advantage of this particular choice of $V$ (Eq.~(\ref{eq3})) is that the Dirac equation $i\partial_t \psi = H \psi$ is analytically solvable.

By solving the Dirac equation, one finds that the lowest energy level $M$, which we shall call {\it constituent quark mass}, and the probability density $\rho \equiv \psi^{\dagger} \psi$ of a single quark in the s-wave state are given by 
\begin{equation}
	M = \sqrt{2\sqrt{(qB/2)^2 + \sigma_{\perp}^2}  + \sigma_{\parallel} +m^2 -qBs }  \label{eq4}
\end{equation}
and
\begin{eqnarray}
	\rho &=& |N|^2 {\rm e}^{-\sqrt{(qB/2)^2 + \sigma_{\perp}^2} r^2 }{\rm e}^{- \sigma_{\parallel} z^2} \nonumber \\
		 &\;\times& \left[ 1 + \frac{ (\sqrt{(qB/2)^2 + \sigma_{\perp}^2}-qBs/2)^2r^2 +   \sigma_{\parallel}^2 z^2}{(M + \sqrt{m^2 + \sigma_{\perp}^2 r^2 + \sigma_{\parallel}^2 z^2})^2} \right].   \label{eq41}
\end{eqnarray}
Here, $s = 1\ {\rm for}\ {\rm spin\ up}~(\uparrow)$ and $-1\ {\rm for}\ {\rm spin\ down}~(\downarrow)$, and $N$ is the normalization constant.  For strong magnetic fields $qB \gg \sigma_{\perp}$, Eq.~(\ref{eq4}) behaves as
\begin{equation}
	M \sim \sqrt{ m^2 + \sigma_{\parallel} + |qB| - qBs }.  \label{eq51}
\end{equation}
The constituent quark mass $M$ increases as $\sqrt{2|qB|}$ for $qs<0$ while $M$ stays almost constant for $qs>0$.  Notice that Eq.~(\ref{eq51}) is independent of the transverse confinement of QCD $\sigma_{\perp} \neq 0$ because hadrons are now strongly squeezed in the transverse direction $\braket{r} \sim 1/\sqrt{qB}$ by the strong magnetic field (see the exponential factor in Eq.~(\ref{eq41})) and the mass contribution from the transverse confinement of QCD $\sim \sigma_{\perp} \braket{r}$ becomes negligible.  We also note that Eq.~(\ref{eq51}) is a slight modification of the naive lowest Landau energy for a charged point-like fermion, $E_{\rm LLL} = \sqrt{m^2+|qB|-qBs}$, due to the longitudinal confinement of QCD $\sigma_{\parallel} \neq 0$.  On the other hand, weak magnetic fields $qB \ll \sigma_{\perp}$ perturbatively deform hadrons to shift their masses as
\begin{equation}
	M \sim M(B=0) - \frac{qBs}{2M(B=0)}.  \label{eq52}
\end{equation}
Here, $M(B=0) = \sqrt{m^2 + 2\sigma_{\perp} + \sigma_{\parallel}}$ is the constituent quark mass at $B=0$.  Equation~(\ref{eq52}) is nothing but the Zeeman splitting formula with the $g$-factor $g=2$.  

\begin{figure}
\begin{center}
\includegraphics[clip, width=0.48\textwidth]{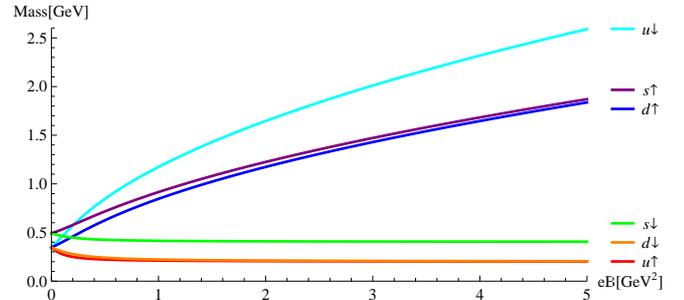}
\caption{\label{fig1}(color online) The constituent quark mass $M$ (Eq.~(\ref{eq4})) of $u\uparrow, u\downarrow, d\uparrow, d\downarrow, s\uparrow, s\downarrow$ as a function of the strength of the magnetic field $eB$.  Parameters are set as $\sigma_{\perp} = \sigma_{\parallel} = (200\;{\rm MeV})^2$ and $m_u=m_d=0\;{\rm MeV}, m_s=350\;{\rm MeV}$.  }
\end{center}
\end{figure}

The $eB$-dependence of the constituent quark mass $M$ (Eq.~({\ref{eq4}})) of $u\uparrow, u\downarrow, d\uparrow, d\downarrow, s\uparrow, s\downarrow$ is plotted in Fig.~\ref{fig1}.  We have set $\sigma \equiv \sigma_{\parallel} = \sigma_{\perp} = (200\;{\rm MeV})^2 \sim \Lambda_{\rm QCD}^2$ and $m_u=m_d=0\;{\rm MeV}, m_s=350\;{\rm MeV}$ so as to reproduce the empirical value of the constituent quark masses at $B=0$, i.e., $M_u, M_d \sim 350\;{\rm MeV}$ and $M_s \sim 500\;{\rm MeV}$.  Here, we have assumed that the parameters $\sigma_{\perp},\sigma_{\parallel}$ do not depend on magnetic fields and thus they are always spherically symmetric $\sigma_{\perp} = \sigma_{\parallel}$ and constant.  We remark that this simplification does not change our qualitative results, while some studies have suggested that the gluon dynamics could be modified, i.e., the confinement force, which is represented by the parameters $\sigma_{\perp},\sigma_{\parallel}$ in our model, might vary under strong magnetic fields through quark loop corrections \cite{sho02, cla14}.

Figure~\ref{fig1} clearly illustrates the explicit breaking of the ${\rm SU(3)}_{\rm flavor} \otimes {\rm SU(2)}_{\rm spin}$ symmetry of hadrons under strong magnetic fields.  Indeed, we have
\begin{equation}
M_{u\downarrow} > M_{s\uparrow} \sim M_{d\uparrow} > M_{s\downarrow} \gtrsim M_{d\downarrow} \sim M_{u\uparrow}  \label{eq5}
\end{equation}
for strong magnetic fields $eB \gg \sigma \sim \Lambda_{\rm QCD}^2$.  The constituent quark mass of $u \downarrow, s\uparrow, d\uparrow$ increases as $\sqrt{2|qB|}$ because $qs<0$, while it stays almost constant for $s \downarrow, d\downarrow, u\uparrow$ because $qs>0$, as is explained in Eq.~(\ref{eq51}).  Thus, the mass hierarchy $M_{u\downarrow} , M_{s\uparrow} , M_{d\uparrow} > M_{s\downarrow} , M_{d\downarrow} , M_{u\uparrow}$ appears.  For the lighter $qs>0$ quarks $u\uparrow, d\downarrow, s\downarrow$, the splitting $M_{s\downarrow} \gtrsim M_{d\downarrow} \sim M_{u\uparrow}$ arises from the current quark mass difference $m_s \gtrsim m_u=m_d$.  For the heavier $qs<0$ quarks $u \downarrow, d\uparrow, s\uparrow$, the splitting $M_{u \downarrow} > M_{d\uparrow} \sim M_{s\uparrow}$ appears due to the electric charge difference $|q_u| = 2e/3 >  |q_d|=|q_s|=e/3$.  Here, the current quark mass difference becomes unimportant because the constituent quark mass of $qs<0$ quarks is largely determined by the electric charge only $M_{qs<0} \sim \sqrt{2|qB|}$.  On the other hand, for weak magnetic fields $eB \ll \sigma \sim \Lambda_{\rm QCD}^2$, the constituent quark mass of $u \uparrow, d\downarrow, s\downarrow$ decreases as $eB$ increases and its magnitude is slightly larger for $u \uparrow$ than for $d\downarrow, s\downarrow$ (see Eq.~(\ref{eq52})).  This is the consequence of the deformation of hadrons by magnetic fields: Hadrons are squeezed by magnetic fields and therefore the mass contribution from the QCD confinement potential $\sim \sigma |{\bm r}_{\rm qq}|$ decreases.  Since $|{\bm r}_{\rm qq}|$ becomes smaller for larger electric charge $q$, we have a stronger reduction of $M_{u \uparrow}$ than $M_{d\downarrow}, M_{s\downarrow}$.  

\begin{table}[!t]
\begin{center}
\caption{Quark components of hadrons under strong magnetic fields.  }\label{table:1}
  \begin{tabular}{c|c} \hline\hline
    Meson 	&  Quarks \\\hline 
    $\pi^0$	& $u \uparrow\bar{u}\downarrow,\  d \downarrow\bar{d}\uparrow$	\\
	$\pi^+$	& $u \uparrow\bar{d}\downarrow$	\\
	$\pi^-$	& $ d \uparrow\bar{u}\downarrow$	\\
	$\eta$	& $u \uparrow\bar{u}\downarrow,\  d \downarrow\bar{d}\uparrow,\ s\downarrow \bar{s}\uparrow$	\\
	$\eta'$	& $u \uparrow\bar{u}\downarrow,\  d \downarrow\bar{d}\uparrow,\ s\downarrow \bar{s}\uparrow$	\\
	$K^0$	& $d\downarrow \bar{s} \uparrow$ \\
	$\bar{K}^0$	& $s\downarrow \bar{d} \uparrow$ \\
	$K^+$	& $u\uparrow \bar{s} \downarrow$ \\
	$K^-$	& $s\uparrow \bar{u} \downarrow$ \\
	$\rho^0$&$ d\uparrow \bar{d}\uparrow,\ d\downarrow \bar{d}\downarrow $ \\
	$\rho^+$&$u\uparrow \bar{d}\uparrow$ \\
	$\rho^-$&$d\downarrow \bar{u}\downarrow$ \\
	$\omega$&$ d\uparrow \bar{d}\uparrow,\ d\downarrow \bar{d}\downarrow $\\
	$\phi$&$ s\uparrow \bar{s}\uparrow,\ s\downarrow \bar{s}\downarrow $\\
	$K^{*0}$& $d\downarrow \bar{s}\downarrow$\\
	$\bar{K}^{*0}$	& $s\uparrow \bar{d}\uparrow$\\
	$K^{*+}$& $u\uparrow \bar{s}\uparrow$\\
	$K^{*-}$& $s\downarrow \bar{u}\downarrow$\\ \hline\hline
  \end{tabular}
  \hspace*{1mm}
  \begin{tabular}{c|c} \hline\hline
    Baryon	& Quarks	\\\hline 
    $p$		& $u \uparrow u\uparrow d\downarrow$ \\
    $n$		& $u\uparrow d\downarrow d\downarrow$ \\
    $\Lambda$& $u \uparrow d\downarrow s\downarrow$ \\
    $\Sigma^+$& $u \uparrow u\uparrow s\downarrow$ \\
    $\Sigma^0$& $u \uparrow d\downarrow s\downarrow$ \\
    $\Sigma^-$& $d\downarrow d\downarrow s\uparrow$\\
    $\Xi^0 $ & $u\uparrow s\downarrow s\downarrow$ \\
    $\Xi^- $ & $d\downarrow s\downarrow s\uparrow$\\
    $\Delta^{++}$& $u \uparrow u \uparrow u \uparrow$ \\
    $\Delta^+$  & $u\uparrow u \uparrow d\uparrow$ \\
    $\Delta^0$  & $u\downarrow d\downarrow d\downarrow$ \\
    $\Delta^-$  & $d\downarrow d\downarrow d\downarrow$ \\
    $\Sigma^{*+}$ & $u\uparrow u\uparrow s\uparrow$ \\
    $\Sigma^{*0}$  & $u\downarrow d\downarrow s\downarrow$ \\
    $\Sigma^{*-}$  & $d\downarrow d\downarrow s\downarrow$ \\
    $\Xi^{*0}$	   & $u\downarrow s\downarrow s\downarrow$ \\
    $\Xi^{*-}$	   & $d\downarrow s\downarrow s\downarrow$ \\
    $\Omega^-$	 & $s\downarrow s\downarrow s\downarrow$ \\
  \hline\hline
  \end{tabular}
\end{center}
\end{table}

Equation~(\ref{eq5}) is an essential relation to construct the proper quark components of hadrons under strong magnetic fields.  This is summarized in Table~\ref{table:1}.  For example, the quark components of $\rho^+$ meson, which is a composite of $u, d$ quarks and has the total electric charge $Q=1$ and the total angular momentum $J = 1$, is given by $u \uparrow \bar{d}\uparrow$.  Indeed, the other $Q=1, J=1$ states such as $u \downarrow \bar{d}\downarrow$ are heavier than $u \uparrow \bar{d}\uparrow$ because of the mass hierarchy Eq.~({\ref{eq5}}).  Simply speaking, the proper quark components of hadrons under strong magnetic fields are determined by maximizing the number of $qs>0$ quarks, whose constituent quark mass stays almost constant, and by minimizing the sum of the electric charge $\sum_{qs<0} |q|$ of $qs<0$ quarks, whose constituent quark mass increases as $\sqrt{2|qB|}$.  We note that all the hadrons under strong magnetic fields are spin-aligned due to this reconstruction of quark components of hadrons.

Now, we are ready to compute hadron mass $M_{\rm Hadron}$ under strong magnetic fields.  By using the constituent quark mass $M_{\rm quark}$ (Eq.~(\ref{eq4})) and the proper quark components of hadrons displayed in Table~\ref{table:1}, we have
\begin{equation}
	M_{\rm Hadron} = \sum_{{\rm quarks} \in {\rm Hadron}} M_{{\rm quark}}.  \label{eq6}
\end{equation}
Here, we have neglected the quark interactions at short distances, one-gluon exchange potential for example, because its mass contribution is always suppressed by the strong coupling constant $\alpha_S$.  The important point in Eq.~(\ref{eq6}) is that the $eB$-dependence of hadron mass $M_{\rm Hadron}$ is largely determined by the number of $qs<0$ quarks and that $M_{\rm Hadron}$ increases as $\sum_{qs<0} \sqrt{2|qB|}$ for strong magnetic fields $eB \gg \sigma \sim \Lambda_{\rm QCD}^2$.  It should be stressed that the mass formula Eq.~(\ref{eq6}) is appropriate for {\it strong} magnetic fields $eB \gg \sigma \sim \Lambda_{\rm QCD}^2$ because it incorporates all the essential properties (a)-(c).  For {\it weak} magnetic fields $eB \lesssim \sigma \sim \Lambda_{\rm QCD}^2$, the mass formula Eq.~(\ref{eq6}) is not adequate to describe hadron masses precisely because the properties (a)-(c) are not the essence for weak magnetic fields.  However, Eq.~(\ref{eq6}) does roughly reproduce the physical hadron masses even for weak magnetic fields because they are largely determined only by the constituent quark masses at $B=0$, which is why we have set the value of the parameters $\sigma, m$ so as to reproduce the empirical value of the constituent quark mass at $B=0$.  In order to obtain a better description for weak magnetic fields, one needs to take into account some other properties which Eq.~(\ref{eq6}) have neglected: The restoration of the ${\rm SU(3)}_{\rm flavor} \otimes {\rm SU(2)}_{\rm spin}$ symmetry of hadrons, the quark interactions at short distances and Chiral corrections.  The chiral corrections are especially important for describing $\pi$ meson masses under weak magnetic fields.  We leave these improvements for a future work.

\begin{figure*}[!t]
\includegraphics[trim=0 20 0 20, width=1.0\textwidth]{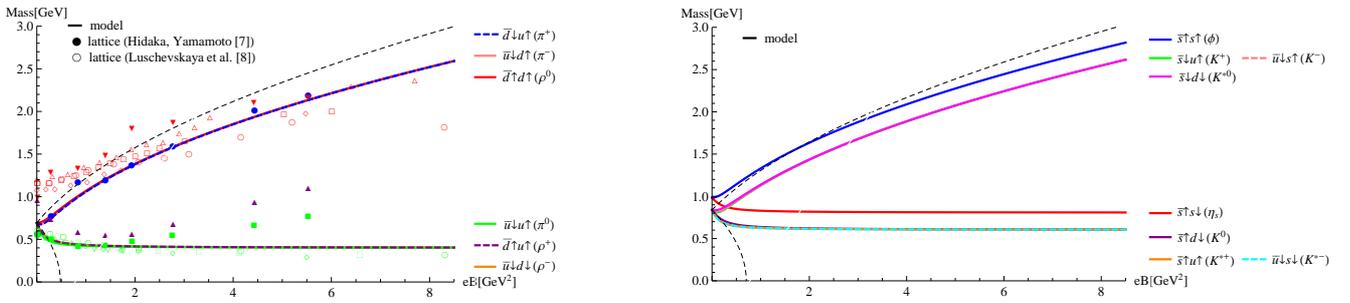}
\caption{\label{fig:meson}(color online) Various meson masses as a function of the strength of the magnetic field $eB$.  [Left] Light mesons.  The thick lines are the model calculation and the symbols are the lattice results; filled symbols for Ref.~\cite{hid13} and open symbols for Ref.~\cite{lus14a}.  Different shapes for open symbols distinguish the lattice space and the volume of the simulations in Ref.~\cite{lus14a}.  The thin black dashed lines are the naive lowest Landau energies for charged $\pi$ mesons, $M_{\pi^{\pm}} = \sqrt{(M_{\pi^{\pm}}(B=0))^2 + eB}$, and charged $\rho$ mesons, $M_{\rho^{\pm}} = \sqrt{(M_{\rho^{\pm}}(B=0))^2 - eB}$.  We note that only $|s_z|=1$ component of $\rho$ meson is considered here because $s_z = 0$ component of $\rho$ meson is mixed up with $\pi$ meson under strong magnetic fields.  [Right] Strange mesons.  The thick lines are the model prediction.  The thin black dashed lines are the naive lowest Landau energies for charged $K$ mesons, $M_{K^{\pm}} = \sqrt{(M_{K^{\pm}}(B=0))^2 + eB}$, and charged $K^*$ mesons, $M_{K^{*\pm}} = \sqrt{(M_{K^{*\pm}}(B=0))^2 - eB}$.  }
\end{figure*}

\begin{figure*}[!t]
\includegraphics[trim=13 20 16 20 , width=1.0\textwidth]{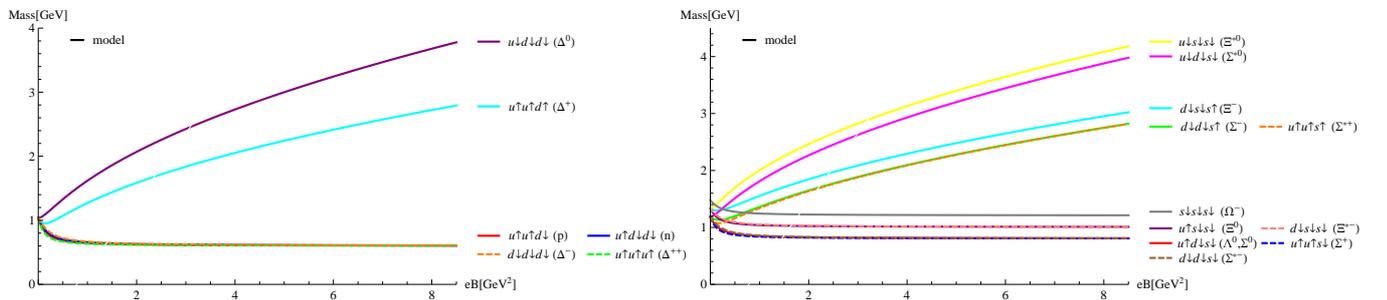}
\caption{\label{fig:baryon}(color online) Various baryon masses as a function of the strength of the magnetic field $eB$.  The thick lines are the model prediction.  [Left] Light baryons.  [Right] Strange baryons.  }
\end{figure*}

\section{Hadron Masses}
 
We analytically calculate various hadrons masses by using Eq.~(\ref{eq6}) and the results are plotted in Figs.~\ref{fig:meson}~and~\ref{fig:baryon}.

The masses of light mesons composed of $u,d$ quarks only are plotted in the left panel of Fig.~\ref{fig:meson}.  The thick lines are our model calculation and the symbols are the existing lattice results \cite{hid13, lus14a}.  One can immediately confirm that our model calculation is qualitatively consistent with the lattice results.  Indeed, we reproduce the mass hierarchy
\begin{equation}
	M_{\rho^0} \sim M_{\pi^{\pm}} > M_{\rho^{\pm}} \sim M_{\pi^0} \label{eq8}
\end{equation}
for strong magnetic fields $eB \gg \sigma \sim \Lambda_{\rm QCD}^2$.  There, $M_{\rho^0}, M_{\pi^{\pm}}$ increase as $\sqrt{2|q_d B|}$ because $\rho^0, \pi^{\pm}$ contain $d\uparrow$ or $\bar{d}\downarrow$ which has $qs<0$, while there is no $qs <0$ quark in $\rho^{\pm}, \pi^0$ and thus $M_{\rho^{\pm}}, M_{\pi^0}$ stay almost constant, i.e., have weak $eB$-dependence\footnote{We note that there is a discrepancy in the existing lattice results \cite{hid13, lus14a} on the $eB$-dependence of $M_{\rho^{\pm}}, M_{\pi^0}$ under strong magnetic fields.  Although both studies \cite{hid13, lus14a} reveal that the $eB$-dependence of $M_{\rho^{\pm}}, M_{\pi^0}$ are weak, Ref.~\cite{hid13} claims that $M_{\rho^{\pm}}, M_{\pi^0}$ slowly increase as the magnetic fields get stronger while Ref.~\cite{lus14a} shows there is no such increase in $M_{\pi^0}$ ($M_{\rho^{\pm}}$ is not studied in Ref.~\cite{lus14a}), i.e., $M_{\pi^0}$ stays almost constant.  }.  For weak magnetic fields $eB \ll \sigma \sim \Lambda_{\rm QCD}^2$, $M_{\rho^{\pm}}, M_{\pi^0}$ decrease as $eB$ increases.  This is the consequence of the deformation of hadrons by magnetic fields.  For comparison, we also plotted the naive lowest Landau energies, $M_{\pi^{\pm}} = \sqrt{(M_{\pi^{\pm}}(B=0))^2 + eB}$ and $M_{\rho^{\pm}} = \sqrt{(M_{\rho^{\pm}}(B=0))^2 - eB}$, respectively, for $\pi^{\pm}$ and $\rho^{\pm}$ as point-like particles in the thin black dashed lines.  The deviation of our model from these lines at large $eB$, which is consistent with the lattice results, reflects the importance of the internal quark structure of hadrons.

Now, we turn to the masses of strange mesons composed of $u,d,s$ quarks.  The results are plotted in the right panel of Fig.~\ref{fig:meson}.  Our model predicts
\begin{equation}
M_{\phi} \gtrsim M_{K^{*0}} \sim M_{K^{\pm}} > M_{\eta_s} \gtrsim M_{K^0} \sim M_{K^{*\pm}}
\end{equation}
for strong magnetic fields $eB \gg \sigma \sim \Lambda_{\rm QCD}^2$.  For the major hierarchy $M_{\phi},M_{K^{*0}},M_{K^{\pm}} > M_{\eta_s}, M_{K^0}, M_{K^{*\pm}}$, the interpretation is the same as that for the light meson masses.  The former hadrons contain $qs = -e/3 < 0$ quark but the latter do not.  The minor splittings $M_{\phi} \gtrsim M_{K^{*0}}, M_{K^{\pm}}$ and $M_{\eta_s} \gtrsim M_{K^0}, M_{K^{*\pm}}$ appear because of the current quark mass difference $m_s \gtrsim m_u=m_d$.  We also show the naive lowest Landau energies for $K^{\pm}, K^{*\pm}$ as point-like particles in the thin black dashed lines.  We again observe the deviation between our model and the naive lowest Landau energies due to the internal quark structure of hadrons.

The masses of light baryons (left panel of Fig.~\ref{fig:baryon}) and strange baryons (right panel of Fig.~\ref{fig:baryon}) are also investigated.  The physics is the same as that for the meson masses: The number of $qs<0$ quarks, the sum of the electric charge $\sum_{qs<0} |q| $ of $qs < 0$ quarks and the number of strange quarks determine the mass hierarchy for strong magnetic fields $eB \gg \sigma \sim \Lambda_{\rm QCD}^2$.  Our model predicts mass hierarchies 
\begin{equation}
M_{\Delta^0} > M_{\Delta^+} > M_{\Delta^-} \sim M_{n} \sim M_{p} \sim M_{\Delta^{++}}
\end{equation}
for light baryons and
\begin{eqnarray}
	&&M_{\Xi^{*0}} \gtrsim M_{\Sigma^{*0}} > M_{\Xi^-} \gtrsim M_{\Sigma^-} \sim M_{\Sigma^{*+}} > M_{\Omega^-} \nonumber\\
	&&\ \gtrsim M_{\Xi^{*-}} \sim M_{\Xi^0} \gtrsim M_{\Sigma^{*-}} \sim M_{\Lambda^0} \sim M_{\Sigma^0} \sim M_{\Sigma^+}  
\end{eqnarray}
for strange baryons.

\section{Summary and Discussion}
 
We have studied hadron masses under strong magnetic fields.  We have developed a quark model which incorporates the explicit breaking of the ${\rm SU(3)}_{\rm flavor} \otimes {\rm SU(2)}_{\rm spin}$ symmetry and the strong deformation of hadrons, which are identified to be the essential properties to describe hadron masses under strong magnetic fields.  Various hadron masses, including baryons and hadrons with strangeness, are analytically calculated by the model.  In particular, the $eB$-dependence of $M_{\pi}, M_{\rho}$ are qualitatively consistent with the recent lattice results \cite{hid13, lus14a, lus14b}.  The model also gives us a clear explanation why there is a non-trivial $eB$-dependence in hadron masses under strong magnetic fields: Under strong magnetic fields exceeding the QCD energy scale $eB \gg \Lambda_{\rm QCD}^2$, only the quarks in the lowest Landau level become important.  In the lowest Landau level, the constituent quark mass increases as $\sqrt{2|qB|}$ for $qs<0$ quarks while it stays almost constant for $qs>0$ quarks.  Thus, the $eB$-dependence of hadron masses is largely determined by the sum of constituent quark mass of $qs<0$ quarks inside a hadron.

Since hadron masses are one of the most basic properties of hadrons, the results of this study have a wide range of applications when discussing hadron physics under strong magnetic fields.  Let us illustrate some examples.  One example is decay modes of hadrons: Some decay modes are kinematically suppressed or enhanced because of the mass hierarchy under strong magnetic fields (see Figs.~\ref{fig:meson}~and~\ref{fig:baryon}).  The modification to decay modes was suggested by Ref.~\cite{che10} which discussed a suppression of $\rho$ meson decays.  Our study suggests that other decay modes are also modified, for example, $K^0 \rightarrow \pi^+\pi^-$ is suppressed so that the lifetime of $K^0$ may become longer.  Another example is the equation of state (EoS) of nuclear matter under strong magnetic fields.  This is important for the physics of neutron stars.  Not only the hadron masses but also interactions between hadrons would affect the EoS.  This is because hadrons are spin-aligned under strong magnetic fields as displayed in Table~\ref{table:1} and hence the spin-dependent part of the hadronic interactions would change.

For the further improvement of our model, it may be important to consider the gluon dynamics: The gluon dynamics could be modified under strong magnetic fields through quark loop corrections.  As a result, the confinement force, which is represented by the parameters $\sigma_{\perp}, \sigma_{\parallel}$ in our model, could depend on magnetic fields \cite{sho02, cla14} as already mentioned.  Also, it is discussed that the constituent quark mass $M$ could vary (magnetic catalysis; see chapter 2 of Ref.~\cite{kaz13} for a review).  If this is the case, the constituent quark mass $M$ acquires additional mass contribution $\sim ({\rm small\ number}) \times \sqrt{|qB|}$ under strong magnetic fields $eB \gg \Lambda_{\rm QCD}^2$.  This results in the slow increase of $M_{\rho^+}, M_{\pi^0}$ and the small splitting $M_{\rho^+} \gtrsim M_{\pi^0}$ observed in the lattice study \cite{hid13}.  This splitting $M_{\rho^+} \gtrsim M_{\pi^0}$ is explained by the constituent mass splitting $M_{u\uparrow} = M_{\bar{u}\downarrow}  \gtrsim M_{d\downarrow} = M_{\bar{d}\uparrow}$ due to the electric charge difference $|q_u| > |q_d|$.  In this situation, the ground state of $\pi^0$ would be given by $\pi_{d}^{0} = \bar{d}\uparrow d\downarrow$, not by $\pi_{u}^{0} = \bar{u}\downarrow u\uparrow$.  Thus, we have $ M_{\rho^+} = M_{u\uparrow} + M_{\bar{d}\uparrow} \gtrsim M_{d\downarrow} + M_{\bar{d}\uparrow} = M_{\pi^0}$.  

\section*{Acknowledgements} 
The author thanks H.~Fujii and K.~Itakura for fruitful discussions and comments.

\end{document}